\documentclass[onecolumn]{aa}
\usepackage{graphics,natbib,amssymb}
\usepackage{rotating}
\usepackage{graphicx}
\usepackage{txfonts}
\newcommand{\bm}[1]{\mbox{\boldmath $#1$}}
\newcommand{\bM}[1]{\mbox{\textbf{\textsf{#1}}}}
\begin{document}
   \title{Constraining the solutions of an inverse method of stellar population synthesis}

   \subtitle{}

   \author{J. Moultaka
          \inst{1} \inst{2}
          \and
          C. Boisson\inst{2}
          \and
          M. Joly\inst{2}
          \and 
          D. Pelat\inst{2}}

%   \offprints{G. Wuchterl}

   \institute{Physikalisches Institut, Z\"ulpicher Str. 77, 50937 K\"oln\\
           \email{moultaka@ph1.uni-koeln.de}
         \and
            LUTH, Observatoire de Meudon,
           5, place Jules Janssen,
           92190 Meudon cedex, France \\ 
%             \thanks{The university of heaven temporarily does not
%                     accept e-mails}
             }

   \date{Received ; accepted }

   \abstract{
In three previous papers (Pelat 1997, 1998 and Moultaka \& Pelat
2000), we set out an inverse stellar population synthesis method which
uses a database of stellar spectra. Unlike other methods, this one
provides a full knowledge of all possible solutions as well as a good
estimation of their stability; moreover, it provides the unique
approximate solution, when the problem is overdetermined, using a
rigorous minimization procedure. In Boisson et al. (2000), this method
has been applied to 10 active and 2 normal galaxies.\\
In this paper we analyse the results of the method after constraining
the solutions. Adding {\it a priori} physical conditions on the
solutions constitutes a good way to regularize the synthesis
problem. As an illustration we introduce physical constraints on the
relative number of stars taking into account our present knowledge of
the initial mass function in galaxies. In order to avoid biases on the
solutions due to such constraints, we use constraints involving only
inequalities between the number of stars, after dividing the H-R
diagram into various groups of stellar masses.\\
We discuss the results for a well-known globular cluster of the
   galaxy M31 and discuss some of the galaxies studied in Boisson et
   al. (2000). We find that, given the spectral resolution and the
   spectral domain, the method is very stable according to such
   constraints (i.e. the constrained solutions are almost the same as
   the unconstrained one). However, an additional information can be
   derived about the evolutionary stage of the last burst of star
   formation, but the precise age of this particular burst seems to be
   questionable.   }

\titlerunning 
{Constraining the solutions of a stellar population synthesis method}
   \maketitle
%
%________________________________________________________________

\section{Introduction}\label{sec:intro}

The search of the stellar populations inside unresolved galaxies has
been the aim of several studies since the sixtees. Two different
approaches have been adopted for this purpose: the direct approach of
which methods are usually called ``the evolutive synthesis methods''
(e.g. Tinsley 1972, Charlot \& Bruzual 1991, Bruzual \& Charlot 1993,
Leitherer et al. 1999, Fioc \& Rocca-Volmerange 1997, Vazdekis, 1999, Bruzual \& Charlot 2003) and the
inverse one of which these are called ``the synthesis methods''
(e.g. Faber 1972, O'Connell 1976, Joly 1974, Bica 1988, Schmidt et
al. 1989, Silva 1991, Pelat 1997,1998, Moultaka \& Pelat 2000). \\ In
the first approach, one decides an {\it a priori} model for the
history of the star formation occuring inside the studied galaxy, and
by means of theoretical stellar evolutionary tracks and of a stellar
database, derives quantities that are directly compared to the
observed ones. From different input models, one retains the model that
best fits the observed quantities.\\ 

In the inverse approach, usually, no {\it a priori} model is necessary
to derive the stellar populations and one uses exclusively the
observables in order to deduce the stellar spectral types and
luminosity classes by means of a minimization procedure. This
minimization task is not a very simple one since the absolute minimum
is usually difficult to find because the ``objective function'' (which
is the function that one has to minimize) can rarely be minimized
analytically.\\

Whatever the approach is, the problem of stellar population synthesis
often suffers from the lack of true solutions or, on the contrary,
from their degeneracy (i.e. multiple solutions) and/or finally from
their instability (i.e. small errors around the observations can
induce discontinuities in the solutions). These three inconveniences
have been controlled in the inverse method described in Pelat
(1997,1998, hereafter Paper I and II) where all the solutions are well
identified and the minimization procedure is rigorously treated, as
well as in Moultaka \& Pelat (2000) where the stability of the various
solutions is analysed (hereafter Paper III). \\

In this paper, we study the influence upon the different solutions of
astrophysical constraints included {\it a priori} when searching for a
solution. The concept of constraining the solutions in the inverse
methods, has been adopted by O'connell (1976), Pickles (1985) and
Silva (1991) but no estimation of the induced bias has been given by
these authors to our knowledge.\\
   
In the next section, we recall briefly the inverse method and its
error analysis; in the third section, we describe constrained
models. Finally, in section \ref{sec:results}, we show and discuss
constrained versus unconstrained results for the globular cluster G170
located in M31, the LINER NGC4278, the starburst NGC3310 and the
Seyfert 2 galaxy NGC2110. In the last section we make a general
description of the behaviour of constrained solutions in the 27
central regions of the twelve galaxies studied in Boisson et
al. (2000), hereafter Paper IV.

%__________________________________________________________________

\section{Description of the inverse method}\label{sec:methode}

As described in Pelat (1997,1998), the present inverse method uses the equivalent widths of galactic spectra absorption lines as observables to be fitted by a combination of the continua fluxes and equivalent widths of a stellar lines database. The basic equation providing the synthetic equivalent widths is the following:
\begin{equation}
W_{\mbox{syn}j}=\frac{\sum_{i=1}^{n_\star} k_{\lambda_0i} W_{ji}I_{ji}}{\sum_{i=1}^{n_\star} k_{\lambda_0i}I_{ji}}\,\,\,\,\,\,\,\mbox{for}\,\,j=1,...,n_{\lambda}
\label{eqsynt}
\end{equation}

where $W_{\mbox{syn}j}$ is the synthetic equivalent width of a line at
wavelength $\lambda_j$, $W_{ji}$ and $I_{ji}$ are respectively, the
equivalent widths and continua fluxes (normalized at a reference
wavelength $\lambda_0$) of the same lines measured in stars of class
$i$; $k_{i}$ is the contribution of star class $i$ to luminosity at
the reference wavelength $\lambda_0$; $n_{\star}$ and $n_{\lambda}$
are respectively the total number of stars considered in the database
and the total number of lines measured in the spectra.\\ To this
equation, we add two physical conditions, that all the stellar
contributions to luminosity $k_{i}$ are positive and their sum equals
one:

\begin{equation}
\bm k \ge \bm 0
\label{kpositif}
\end{equation} 

\begin{equation}
\sum_{i=1}^{n_{\star}}k_i=1
\label{bk=1}
\end{equation}

Thus, having the observed set of galactic equivalent widths $W_{\mbox{obs}j}$, one searchs for the stellar contributions $k_{i}$ satisfying the previous physical conditions and minimizing the following objective function which is the square of what is called the synthetic distance $D$. The synthetic distance represents a kind of $\chi^2$ considering the
$W_{obs\,j}$ as data. In fact $D^2$ would be exactly a $\chi^2$ if $P_j^{-1}$ were
choosen as the $W_{obs\,j}$ variances (i.e. $P_j=\sigma_{W_{obs\,j}}^{-2}$, see Sect. 5 of Paper I for a discussion on $\sigma_{W_{obs\,j}}$ value) :
\begin{equation}
D^2=\sum_{j=1}^{n_{\lambda}}(W_{\mbox{obs}j}-W_{\mbox{syn}j})^2P_j,\,\,P_j\geq 0
\label{D2P}
\end{equation}
In this definition, $P_j$ is a weight characterizing the quality of the equivalent widths measurements. One can eliminate this weight by applying the change of variables $W_j \leftrightarrow W_j \sqrt P_j$. In the following, we will consider this change of variables already made.  
As stated, the problem can either be overdetermined (i.e. there are more equations than unknowns) or underdetermined (i.e there are more unknowns than equations). In the first case, one finds at most a unique solution and, because of observational errors, there is usually not any exact solution; then the adopted solution is the approximate one which minimizes the synthetic distance. In Paper I, it is shown that this minimum is unique and near the true one; in addition, it is demonstrated that when signal to noise ratio goes to infinity, the unique minimum is exactly the true one (see Paper I).\\
In the underdetermined case, one gets an infinite number of solutions which are convex combinations of particular solutions called the extreme solutions (see Paper II).\\

Finally, the search of the error regions around the solutions has been
made in Paper III, giving thus a relevance for the solutions and a
criterium to sort the various solutions in the underdetermined case by
order of merit. An ``ideal'' database (i.e. a database with an infinite spectral resolution, the largest wavelength domain and adapted for the velocity dispersion observed in the studied galaxy) is not degenerate. Then the analysis made in Paper III gives equivalently a condition to the
database (depending on the quality of the observations) allowing to
get a well-defined solution. Indeed, depending on the quality of the observations (i.e. on the size of the galactic error region), the stellar database may become degenerate. As the S/N ratios of the stellar spectra are usually higher than those of the galaxy, more than one star may lay inside the galactic error region in the equivalent widths vector space. This situation leads to very badly determined solutions, because the different stars can not all be distinguished in this case. One has to eliminate such correlated stars from the database in order to obtain a well-defined solution.
In the Appendix, the computation of the
standard deviation around the synthetic distance is described.\\

\section{The constraints model:}\label{sec:contr}

\subsection{The stellar database:}\label{sec:database}
The database used comes from three different stellar libraries Serote Roos et al. (1996), Silva \& Cornell (1992) and Fluks et al. (1994). The spectral resolution of the resulting stellar database is of $11$\AA\ and the spectral domain goes from 5000\AA\ to $8800\AA$. Given the galaxy velocity dispersion, the velocity broadening of the lines is equivalent to the spectral resolution of the stellar database. Thus no correction for velocity dispersion has to be applied. In this domain and at this spectral resolution, we have selected 47 line features and measured their equivalent widths (see Paper IV), some wavelength intervals have not been included in the database because of atmospheric absorption bands not corrected in Silva \& Cornell library. The table of the line intervals is shown in Boisson et al. (2000). Stars included in the database have been chosen so that the H-R diagram is best represented in spectral types and luminosity classes for two metallicities (solar and about twice solar).  

\subsection{The model:}
In general, the synthesis problem is, what is called in mathematical terms, an ``Ill-posed'' problem. This means that the problem may have no solution or a large number of solutions and/or that the solution is not stable against small deviations around the observation. The usual way to overcome this difficulty is to regularize the problem. This can be done by searching the solution of maximal entropy for example or using any other reasonable criterium. We suggest here to regularize the problem using physical criteria in order to obtain a unique and stable solution. This regularization procedure has already been adopted in the previous papers when the positivity conditions ($k_i \ge 0$) were considered. In the present paper, we introduce more physical conditions in order to constrain the solution.\\

 Technically, constraining the solutions of an inverse problem comes to reduce the volume of the simplex of solutions or the synthetic surface as defined in Paper I (which is the set of exact solutions). The problem of such a procedure, is that the solution may be biased by the constraints model while, 
according to the philosophy of the inverse approach, it should not be altered by any {\it a priori} overconstrained model (otherwise it will reflect the model itself and will not provide additional information on the real stellar population). This inconvenience can be checked if the solutions are found to be on the border of the simplex (where this one has been reduced), because in this case, this would mean that the solution has been strongly constrained in order to lay in the reduced simplex (otherwise, it would have layed inside or on the other borders of the simplex).\\
Then we could summarize the process for the search of a solution with the following scheme:\\
Having the data (the spectra), one uses a model in order to derive the stellar contributions to luminosity $k_i$, this is the inverse approach.
 This process could, as shown in Paper II, provide an infinite number of exact solutions (in the underdetermined case) or an approximate solution obtained by a minimization procedure (in the overdetermined case). The obtained solutions are ``mathematical'' solutions which solve rigourously the mathematical problem. They could be stable or unstable against small deviations around the observation. In order to reduce the number of solutions and to insure that the solutions satisfy the physical conditions of the studied object, it is necessary to constrain the problem.\\

Thus, the synthesis problem may be stated as minimizing the synthetic distance of equation (\ref{D2P}) also written in the following form: 
\begin{equation}
D^2=\sum_{j=1}^{n_{\lambda}}\Bigg({\frac{\sum_{i=1}^{n_\star}(W_{\mbox{obs}j}-W_{ji})I_{ji}k_i}{\sum_{i=1}^{n_\star}I_{ji}k_i}}\Bigg)^2
\label{distancesynt}
\end{equation}
(where the change of variable $W_j \leftrightarrow W_j \sqrt P_j$ has been applied).
The set of stellar contributions $\bm{k}~=~(k_1,k_2,...k_{n_{\star}})$ is submitted to the conditions:
\begin{equation}
\bm l\le
\left\{
\begin{array}{c}
\bm k\\
\bM C\bm k
\end{array}\right\}\le \bm u
\label{contr}
\end{equation}
In the previous inequalities, $\bm l$ and $\bm u$ are respectively, lower and upper constant limits and the first inequality takes into account that all contributions are positive and less than one because of condition (\ref{bk=1}), therefore, the first $n_{\star}$ components of vectors $\bm l$ and $\bm u$ are, respectively, equal to zero and one. $\bM C$ is the constraints matrix of $n_{\star}$ columns in which the first line is a vector of components equal to one in order to express constraint number (\ref{bk=1}) then the component number $n_{\star}+1$ of vectors $\bm l$ and $\bm u$ is equal to one. \\

Since the expression of the synthetic distance of equation (\ref{distancesynt}) is not linear in $\bm k$, we decided to minimize instead the linear function $\|{\bM A\bm k}\|^2$ submitted to the above constraints (where $\bM A$ is a matrix defined in Paper I of whose components are $A_{ji}=(W_{obsj}-W_{ji})I_{ji}$). This function happens to be the ``principal synthetic distance'' defined in Paper I as:
\begin{equation}
D^2_{\mbox{prin}}=\sum_{j=1}^{n_{\lambda}}(W_{\mbox{obs}j}-W_{\mbox{syn}j})^2 I_{\mbox{syn}j}^2,\,\,\,\,I_{\mbox{syn}j}=\sum_{i=1}^{n_\star}I_{ji}k_i
\label{D2Isyn}
\end{equation}

As it is stated in Paper I, the problem possesses then a unique solution. Thus, using the constrained least square method, we can derive the unique set of stellar contributions minimizing the principal synthetic distance and satisfying the model constraints (\ref{contr}). Once the principal solution is at hand, we can find the ``principal geometrical'' one which is the nearest solution to the constrained principal solution and minimizes the initial synthetic distance of equation (\ref{distancesynt}). For this purpose, we use the same iterative method as in Paper I where the iteration is made on the synthetic surface. At each step of iteration (m+1) and using the constrained least square method, we search for the new solution $\bm k^{(m+1)}$ that minimises the following function:
\begin{equation}
\parallel\bM X^{(m+1)}\bm k^{(m+1)}-\bm b\parallel^2
\end{equation}
submitted to the constraints of equation \ref{contr} and where $\bm b$ is defined by the relation $\bm b=\bm W_{obs}-\bm W_{syn}+\bM X\bm k^{(m)}$. \\

The solution of the constrained least square method is obtained using procedures from the NAG library.  

\subsection{The constraints:}\label{constraints}

As the spectral resolution and the wavelength range are limited, the
number of uncorrelated stars is also limited. This leads to inherent
incompleteness of the database which together with different
photometric accuracy between galactic and stellar spectra can lead to
non-physical solutions. So, one may have to constrain the solutions, in particular, in such a way that the Initial Mass function of stars (IMF) satisfies the known shape of this function derived from observations of resolved objects. 
\\

\begin{table*}[htbp]
\small
\begin{center}
\begin{tabular}{|c|c|c|}
\hline
Mass group    & Mass interval  & stars\\ 
\hline
1     & $17 M_\odot-30 M_\odot$ & O7-B0V \\
      &  & M2Ia \\
\hline
2     & $ 3M_\odot-17 M_\odot$ &  B3-4V\\
      &  & G0Iab,K4Iab,rG2Iab,rK0II,rK3Iab \\
\hline
3     & $ 1.6M_\odot-3 M_\odot$ &  A1-3V\\
\hline
4     & $ 0.8M_\odot-1.6 M_\odot$ &  F2V,F8-9V,G4V,rG0IV,rG5IV\\
      &  & G0-4III,wG8III,G9III,K4III,M0.5III,M4III,M5III,rG9III,rK3III,rK3III(bis),rK5III \\
\hline
5     & $ \leq 0.8 M_\odot$ & G9-K0V,K5V,M2V,rK0V,rK3V,rM1V \\
\hline
\end{tabular}
\end{center}
\caption{The mass groups cutout of the stellar database in the case of the ``Standard mode''.}
\label{massgroupstd}
\end{table*}

\begin{table*}[htbp]
\small
\begin{center}
\begin{tabular}{|c|c|c|}
\hline
Mass group    & Mass interval  & stars\\ 
\hline
1     & $17 M_\odot-30 M_\odot$ & O7-B0V \\
      &  & M2Ia \\
\hline
2     & $ 3M_\odot-17 M_\odot$ &  B3-4V\\
      &  & G0Iab,K4Iab,rG2Iab,rK0II,rK3Iab \\
\hline
3     & $ 1.6M_\odot-3 M_\odot$ &  A1-3V\\
\hline
4     & $ 1.4M_\odot-1.6 M_\odot$ &  F2V\\
      &  & M4III,M5III\\
\hline
5     & $ 1.1M_\odot-1.4 M_\odot$ & F8-9V,rG0IV,rG5IV\\
      &  & G0-4III,wG8III,G9III,K4III,M0.5III,rG9III,rK3III,rK3III(bis),rK5III \\
\hline
6     & $ 0.8M_\odot-1.1 M_\odot$ & G4V\\
\hline
7     & $  0.7M_\odot-0.8 M_\odot$ & G9-K0V,rK0V,rK3V \\
\hline
8     & $  0.5M_\odot-0.7 M_\odot$ & K5V\\
\hline
9     & $  \leq 0.5 M_\odot$ & M2V,rM1V \\
\hline
\end{tabular}
\end{center}
\caption{The mass groups cutout of the stellar database in the case of the ``Decreasing IMF mode''.}
\label{massgroupimf}
\end{table*}

We first define groups of main
sequence (MS) stars following their range of lifetime. This translates into 
stellar mass intervals. In each of these groups, we include the
evolved stars of the same initial mass. Tables \ref{massgroupstd} and
\ref{massgroupimf} show the cutout of the stellar database in ``mass
groups'' for two different mass resolutions. The mass interval
associated to each star of the base was determined using Schmidt-Kaler
tables (1982). When necessary, we
interpolated the available masses in the tables.  Concerning the
evolved stars, we used the evolutionary tracks of Padova and Geneva
groups to associate an interval of initial masses for the stars. Each star
thus represents an evolutionary stage in a given mass group. We aim in
any case at not privileging one of the classical IMFs (as the
functions of Salpeter 1955, Scalo 1986 and Kroupa et al. 1993).  \\

We will discuss here two different modes of constraints corresponding to the two different
samplings in mass of the H-R diagram shown in tables \ref{massgroupstd} and
\ref{massgroupimf}:\\

{\bf The ``Standard mode'':}\\

In this mode, we impose a general constraint on the number of born stars of
the different mass groups:

\begin{equation}
N_i \leq N_j
\label{standgen}
\end{equation}

where $N_i$ and $N_j$ are respectively the numbers of born stars with
masses located in the mass intervals ]$M_{ia}$,$M_{ib}$[ and
]$M_{ja}$,$M_{jb}$[ and where $M_{jb}\leq M_{ia}$.  Even though a
galaxy is not described by a single burst of star formation, equation
\ref{standgen} is valid as well for all the Main-Sequence (MS) stars
produced by a continuous star formation scenario or by multiple bursts
of star formation.  As stars of higher masses evolve faster than the
ones of smaller masses, equation \ref{standgen} is also valid for all
stars (MS and evolved) belonging to the same mass groups.

This translates into the following constraints set:

\begin{equation}
N_{MS\,i} \leq N_{MS\,j}
\label{standMS}
\end{equation}
\begin{equation}
N_{T\,i}\leq N_{T\,j}
%N_{MS\,i}+N_{evolved\,i} \leq N_{MS\,j}+N_{evolved\,j}
\label{standtot}
\end{equation}

where $N_{MS\,i}$ and $N_{T\,i}$ are respectively the number of MS
stars and the total number of stars (MS and evolved stars) in a given
mass interval.  In table \ref{massgroupstd}, the mass intervals are
chosen the narrowest possible such that equations \ref{standMS} and
\ref{standtot} are satisfied in the case of the classical IMFs.

A consequence of introducing such constraints is that the resulting solutions will present higher synthetic distances than in the unconstrained solution, and therefore the fit will be less good.\\

{\bf The ``Decreasing IMF mode'':}\\

The second mode of constraints called ``the Decreasing IMF mode''
states that all initial mass functions are decreasing
functions. Consequently, for a burst of star formation, the total
number of born stars $N_{Ti}$ of a given mass group 
($N_{Ti}=N_{MS\,i}+N_{evolved\,i}$) of a given group of stars
corresponding to a mass interval ]$M_{ia}$,$M_{ib}$[ with a mean mass
$M_i$ ($M_i=\frac{M_{ia}+M_{ib}}{2}$) is such that:

\begin{equation}
N_{Ti} \leq \frac{\Delta{\log M_i}*N_{Tj}}{\Delta{\log M_j}}
\label{imfdec}
\end{equation}

where $M_j < M_i$. \\ 

$\Delta{\log M_i}$ and $\Delta{\log
M_j}$ are respectively the logarithmic lengths of each mass
interval. The difference with respect to equation 8 is that the number of stars
in each group is weighted by the mass.

As in the ``Standard mode'', equation \ref{imfdec} is valid for the
number of MS stars on the one hand and the total number (MS and
evolved) stars on the other hand. Then, one can write:

\begin{equation}
N_{MS\,i} \leq \frac{\Delta{\log M_i}*N_{MS\,j}}{\Delta{\log M_j}}
\label{imfdecMS}
\end{equation}
\begin{equation}
N_{T\,i} \leq \frac{\Delta{\log M_i}*N_{T\,j}}{\Delta{\log M_j}}
\label{imfdectot}
\end{equation}

where the notations are the same as for the ``Standard mode''.\\

As equation (\ref{imfdec}) depends on the mass intervals, the resulting
mass cutout is more refined for the low mass stars (see table
\ref{massgroupimf}).  Because of the better resolution of the mass
cutout in this case, the number of constraints is increased compared
to the ``Standard mode''. Consequently, the synthetic distance in this
mode of constraints will be higher.

The previous description of the adopted constraints shows that these take the form of large inequalities between the numbers of stars or equivalently, in the optical domain, between the stellar contributions to luminosity at the reference wavelength $k_{\lambda_{0i}}$. The relation between the number of stars of class $i$ and their contribution to luminosity is as follows:
\begin{equation}
n_{i}=\frac{k_{\lambda_{0i}} L_{\lambda_{0}{\mbox{gal}}}}{l_{\lambda_{0i}}}
\label{nbrstar}
\end{equation} 
where $ L_{\lambda_{0}{\mbox{gal}}}$ is the galactic (or the cluster) luminosity at the reference wavelength $\lambda_0$ and $l_{\lambda_{0i}}$ is the luminosity of a star of class $i$ at the same wavelength.\\
As the constraints have the form of large inequalities
(i.e. equalities are allowed) one may obtain solutions on the border of
the domain of constraints (i.e. if the constraints are satisfied with
equalities), the solution is probably over-constrained. In such a
case, one has to conclude that the constraint set is not "appropriate"
to the data.

%fin

%===================================================================

\section{Results}\label{sec:results}

The effect of adding constraints to the inverse method has been tested
on the globular cluster G170 located in M31 and three central regions
of galaxies corresponding to three types of galaxy activity considered
in Paper IV.

 As the aim of the present paper is to study the effects of
constraining the solutions, we will only concentrate on a description
and on the comparison of the different solutions. Therefore no
complete physical discussion (especially in the case of the galaxies)
will be made in the following sections since astrophysical issues
about these objects have been plenty discussed in Paper IV (for a full
discussion, we invite the reader to refer to the paper).\\

In tables \ref{G170} to \ref{NGC2110}, we show the list of stars
present in the database (column one), the unconstrained solution (only
statisfying the positivity condition (\ref{kpositif}) and equation
(\ref{bk=1})) in column 2. In the two following columns, we show the
mass intervals corresponding to the mass cutout for the ``Decreasing IMF mode'' and the constrained solutions of this mode. Columns 5 and 6 exhibit the mass intervals and the constrained solutions in the case of the ``Standard mode''. Finally, in some examples, we show particular solutions
where we imposed a fixed value to one or more stellar contributions.\\
The solutions  list the stellar contributions,  $\bm k_i$, to
luminosity  at the reference wavelength $\lambda_0 = 5450
\AA$.

\subsection{The stellar cluster G170}
In order to validate our method, we compare the constrained and
unconstrained solutions obtained for a single stellar population
system, namely, the well studied globular cluster G170 located in the
galaxy M31. The globular cluster spectrum is taken from Jablonka et al
(1992).

The unconstrained solution (see table \ref{G170}) shows that this
cluster has a solar metallicity. The turnoff in G4V suggests an age of
about $10^{10}$ years and the contribution of dwarf stars to the luminosity of $57\%$
 shows that the luminosity in the optical is dominated
by the MS stars. The metallicity is clearly solar (the contribution of stars of solar metallicity is about $\sim92\%$). The measured internal reddening E(B-V) is of $\sim
0.05$. This value of the reddening is determined as the correction to
be applied to the observed spectrum to match the synthetic one. In
this process the Galactic law is parametrized as in Howarth (1983).

These results agree well with those found by Jablonka et al. (1992); as a
matter of fact, the authors concluded, using measurements of
equivalent widths of absorption lines, that this cluster has a solar
metallicity as well as an age of $2\,10^{10}$ years.

 The ``Decreasing IMF mode'' presents a solution with acceptable
 synthetic distance in the sense that its value is at $\sim 1
 \sigma$ from the synthetic distance value of the unconstrained
 solution, but the contributions are slightly different from the
 latter: only $\sim 38\%$ of the optical luminosity is
 due to MS stars and $\sim20\%$ of it is due to metallic stars.

The constrained solution of the ``Standard mode'' is equal to the unconstrained one. This result shows that the latter is physically acceptable.

The difference between the two solutions, ``Decreasing IMF mode'' and ``Standard'' or unconstrained mode is illustrated in Fig.1.

\begin{figure*}
\resizebox{15cm}{!}{{\includegraphics{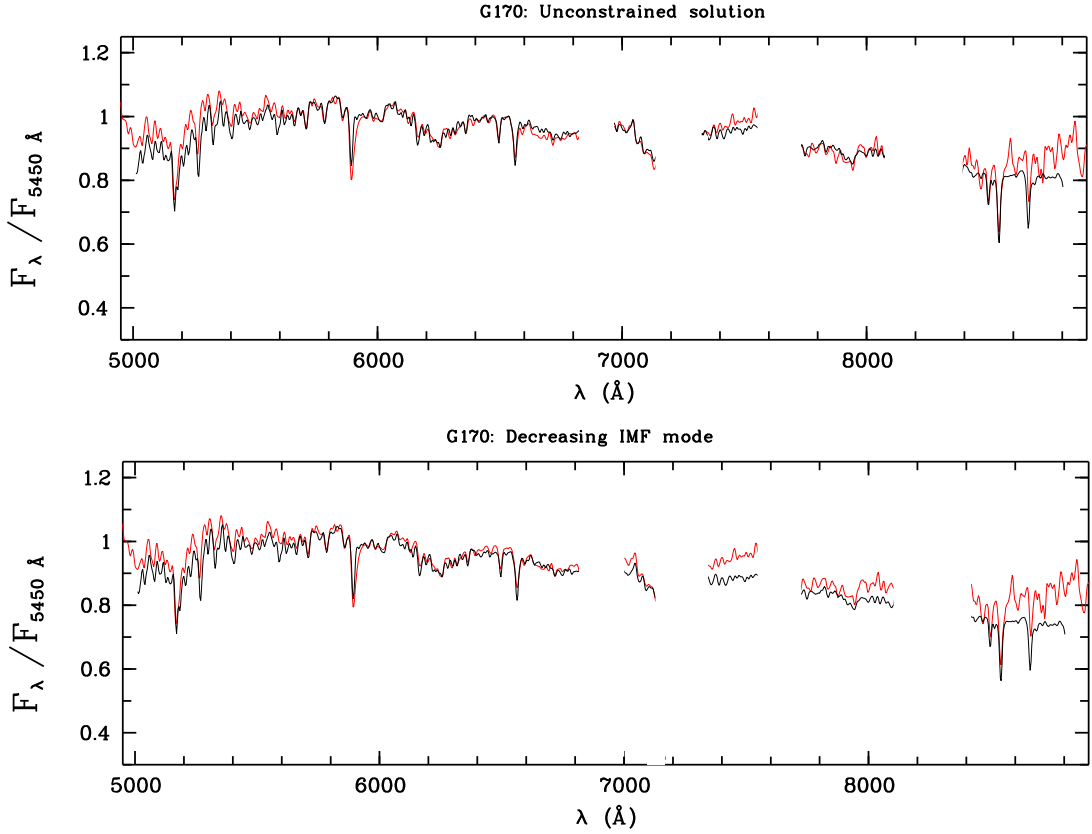}}}
\caption{NGC G170: Synthetic and observed spectra respectively in dark and light lines for different solutions.}
\label{G170a.eps}
\end{figure*}

%\begin{figure*}
%\resizebox{15cm}{!}{{\includegraphics{G170b.eps}}}
%\caption{NGC G170: Same as Fig. \ref{G170a.eps}}
%\label{G170b.eps}
%\end{figure*}

%\begin{figure*}
%\resizebox{18cm}{!}{{\includegraphics{HRG170_1ter.eps}}}
%\caption{The globular cluster G170: HR diagram of the different solutions for which spectra are plotted in Fig. \ref{G170a.eps}. The size of the squares is proportional to the stellar contributions $k_i$ and the error bars represent the standard deviations around the $k_i$ listed in table \ref{G170}. The open and filled squares represent respectively the solar and over-abundant stars.}
%\label{HRG170_1.eps}
%\end{figure*}

%\begin{figure*}
%\resizebox{6cm}{!}{{\includegraphics{HRG170_2ter.eps}}}
%\caption{The globular cluster G170: HR diagram of the particular solution (2).}
%\label{HRG170_2.eps}
%\end{figure*}

%\begin{sidewaystable*}[htbp]
\begin{table*}
\small
\begin{center}
\begin{tabular}{|r|c|c|c|c|c|} %c|c|c|c|c|c|}
\hline
          & Unconstrained      & Mass interval & Dec. IMF     & Mass interval & Standard    \\%& Standard    & Standard     %& Particular  & Particular\\ 
Star      & solution  & Dec. IMF mode &  mode    &  Standard mode &  mode         \\%&& Type 1      & Type 2      & Type 3        & solution (1)  & solution (2)\\ 
\hline
O7-BOV    & $0$          & $17 M_\odot-30 M_\odot$ & $0$         & $17 M_\odot-30 M_\odot$ &$0$         \\%& $0$         & $0$           & $0.1\pm7$  & \underline{$7$}\\
B3-4V     & $0$          & $ 3M_\odot-17 M_\odot$ & $0.01\pm2$  & $ 3M_\odot-17 M_\odot$ & $0$         \\%& $0$         & $0$           & \underline{$5$}  & \underline{$5$}\\
A1-3V     & $0$          & $ 1.6M_\odot-3 M_\odot$ & $0$         & $ 1.6M_\odot-3 M_\odot$ & $0$         \\%& $0$         & $0$           & $0.2\pm9$  & $0.2\pm4$\\ 
F2V       & $0$          & $ 0.8M_\odot-1.6 M_\odot$ &$0$         & $ 1.4M_\odot-1.6 M_\odot$ &$0$         \\%& $0$         & $4\pm7.5$     & $0.02\pm15$  & $0$\\
F8-9V     & $0$          & $ 0.8M_\odot-1.6 M_\odot$ &$0$         & $ 1.1M_\odot-1.4 M_\odot$ &$0$         \\%& $0$         & $0$           & $2\pm2$  & $0$\\
G4V       & $22\pm11$    & $ 0.8M_\odot-1.6 M_\odot$ &$7\pm12$    & $ 0.8M_\odot-1.1 M_\odot$ &$22\pm11$   \\%& $22\pm11$   & $0$           & $6\pm10$  & $5\pm10$\\ 
G9-K0V    & $0$          & $ \leq 0.8 M_\odot$ &$0$         & $  0.7M_\odot-0.8 M_\odot$ &$0$         \\%& $0$         & $0$           & $0$  & $0$\\
K5V       & $27\pm7$     & $ \leq 0.8 M_\odot$ &$11\pm5$    & $  0.5M_\odot-0.7 M_\odot$ &$27\pm7$    \\%& $27\pm7$    & $20.5\pm5$    & $13\pm19$  & $15\pm0.01$\\ 
M2V       & $0$          & $ \leq 0.8 M_\odot$ &$2\pm1$     & $  \leq 0.5 M_\odot$ &$0$         \\%& $0$         & $0$           & $2\pm3$  & $2\pm1$\\ 
rG0IV     & $0$          & $ 0.8M_\odot-1.6 M_\odot$ &$0$         & $ 1.1M_\odot-1.4 M_\odot$ &$0$         \\%& $0$         & $0$           & $0$  & $0$\\  
rG5IV     & $0$          & $ 0.8M_\odot-1.6 M_\odot$ &$0$         & $ 1.1M_\odot-1.4 M_\odot$ &$0$         \\%& $0$         & $0$           & $0$  & $0$\\  
rK0V      & $8\pm6$      & $ \leq 0.8 M_\odot$ &$18\pm5.5$  & $  0.7M_\odot-0.8 M_\odot$ &$8\pm6$     \\%& $8\pm6$     & $16\pm2$      & $20\pm29$  & $23\pm9$\\  
rK3V      & $0$          & $ \leq 0.8 M_\odot$ &$0$         & $  0.7M_\odot-0.8 M_\odot$ &$0$         \\%& $0$         & $0$           & $0$  & $0$\\  
rM1V      & $0$          & $ \leq 0.8 M_\odot$ &$0$         & $  \leq 0.5 M_\odot$ &$0$         \\%& $0$         & $0$           & $0$  & $0$\\  
G0-4III   & $31\pm13$    & $ 0.8M_\odot-1.6 M_\odot$ &$50\pm7$    & $ 1.1M_\odot-1.4 M_\odot$ &$31\pm13$   \\%& $31\pm13$   & $47\pm12$     & $39\pm58$  & $30\pm13$\\  
wG8III    & $0$          & $ 0.8M_\odot-1.6 M_\odot$ &$0.01\pm6$  & $ 1.1M_\odot-1.4 M_\odot$ &$0$         \\%& $0$         & $0$           & $0$  & $0$\\  
G9III     & $0$          & $ 0.8M_\odot-1.6 M_\odot$ &$0$         & $ 1.1M_\odot-1.4 M_\odot$ &$0$         \\%& $0$         & $0$           & $0$  & $0$\\  
K4III     & $0$          & $ 0.8M_\odot-1.6 M_\odot$ &$0$         & $ 1.1M_\odot-1.4 M_\odot$ &$0$         \\%& $0$         & $0$           & $0$  & $0$\\ 
M0.5III   & $11\pm1$     & $ 0.8M_\odot-1.6 M_\odot$ &$8\pm1$     & $ 1.1M_\odot-1.4 M_\odot$ &$11\pm1$    \\%& $11\pm1$    & $11\pm1$      & $8\pm11$  & $9\pm2$\\ 
M4III     & $0$          & $ 0.8M_\odot-1.6 M_\odot$ &$0$         &$ 1.4M_\odot-1.6 M_\odot$ & $0$         \\%& $0$         & $0$           & $0$  & $0$\\ 
M5III     & $1\pm0.1$    & $ 0.8M_\odot-1.6 M_\odot$ &$1\pm0.1$   & $ 1.4M_\odot-1.6 M_\odot$ &$1\pm0.1$   \\%& $1\pm0.1$   & $1\pm0.1$     & $0.8\pm1$  & $1\pm0.2$\\ 
rG9III    & $0$          & $ 0.8M_\odot-1.6 M_\odot$ &$0$         & $ 1.1M_\odot-1.4 M_\odot$ &$0$         \\%& $0$         & $0$           & $0$  & $0$\\ 
rK3III    & $0$          & $ 0.8M_\odot-1.6 M_\odot$ &$0$         & $ 1.1M_\odot-1.4 M_\odot$ &$0$         \\%& $0$         & $0$           & $0.5\pm2$  & $1\pm2$\\ 
rK3IIIbis & $0$          & $ 0.8M_\odot-1.6 M_\odot$ &$0$         & $ 1.1M_\odot-1.4 M_\odot$ &$0$         \\%& $0$         & $0$           & $0$  & $0$\\ 
rK5III    & $0$          & $ 0.8M_\odot-1.6 M_\odot$ & $2\pm2$     & $ 1.1M_\odot-1.4 M_\odot$ &$0$         \\%& $0$         & $0$           & $2\pm4$  & $2\pm1$\\ 
G0Iab     & $0$          & $ 3M_\odot-17 M_\odot$ & $0$         & $ 3M_\odot-17 M_\odot$ &$0$         \\%& $0$         & $0$           & $0$  & $0$\\ 
K4Iab     & $0$          & $ 3M_\odot-17 M_\odot$ & $0$         & $ 3M_\odot-17 M_\odot$ & $0$         \\%& $0$         & $0$           & $0$  & $0$\\ 
M2Ia      & $0$          & $17 M_\odot-30 M_\odot$ & $1\pm1$     & $17 M_\odot-30 M_\odot$ & $0$         \\%& $0$         & $0$           & $0.4\pm0.7$  & $0$\\ 
rG2Iab    & $0$          & $ 3M_\odot-17 M_\odot$ & $0$         & $ 3M_\odot-17 M_\odot$ & $0$         \\%& $0$         & $0$           & $0$  & $0$\\ 
rK0II     & $0$          & $ 3M_\odot-17 M_\odot$ & $0$         & $ 3M_\odot-17 M_\odot$ & $0$         \\%& $0$         & $0$           & $0$  & $0$\\ 
rK3Iab    & $0$          & $ 3M_\odot-17 M_\odot$ & $0$         & $ 3M_\odot-17 M_\odot$ & $0$         \\%& $0$         & $0$           & $0$  & $0$\\ 
\hline
$D^2$ or $\chi^2$    & $7.6\pm0.4$ & & $8.0\pm0.35$ & &  $7.6\pm0.4$ \\%& $7.6\pm0.4$ & $7.7\pm0.4$   & $8.1\pm0.3$  & $8.3\pm0.3$\\ 
\hline
E(B-V)    & $0.05$       & & $0.08$     & & $0.05$     \\%& $0.05$        & $0.08$  & $0.09$ & $0.09$\\
\hline
\end{tabular}
\end{center}
\caption{Results of the spectral synthesis of the globular cluster G170. Each column displays the stellar contributions to luminosity at $\lambda_0=5450\AA$ with their standard deviations for the various solutions (with different modes and types of constraints) as well as for the unconstrained solution. The over-abundant and under-abundant stars are respectively designated by an ``r'' and a ``w'' preceding the spectral type and luminosity class. $D^2$ is the synthetic distance (or the ``mean'' residual EW) with its standard deviation computed in the appendix and E(B-V) is the reddening that can be deduced as described in the text.}
\label{G170}
\end{table*}
%\end{sidewaystable*}

\subsection{The nucleus of the LINER NGC 4278}
As shown in table \ref{NGC4278} column 2, the unconstrained model suggests that the optical spectrum of this region is dominated by dwarf stars as well as a metal rich population. An internal reddening E(B-V) of $\sim 0.02$ is detected. 

\begin{figure*}
\resizebox{15cm}{!}{{\includegraphics{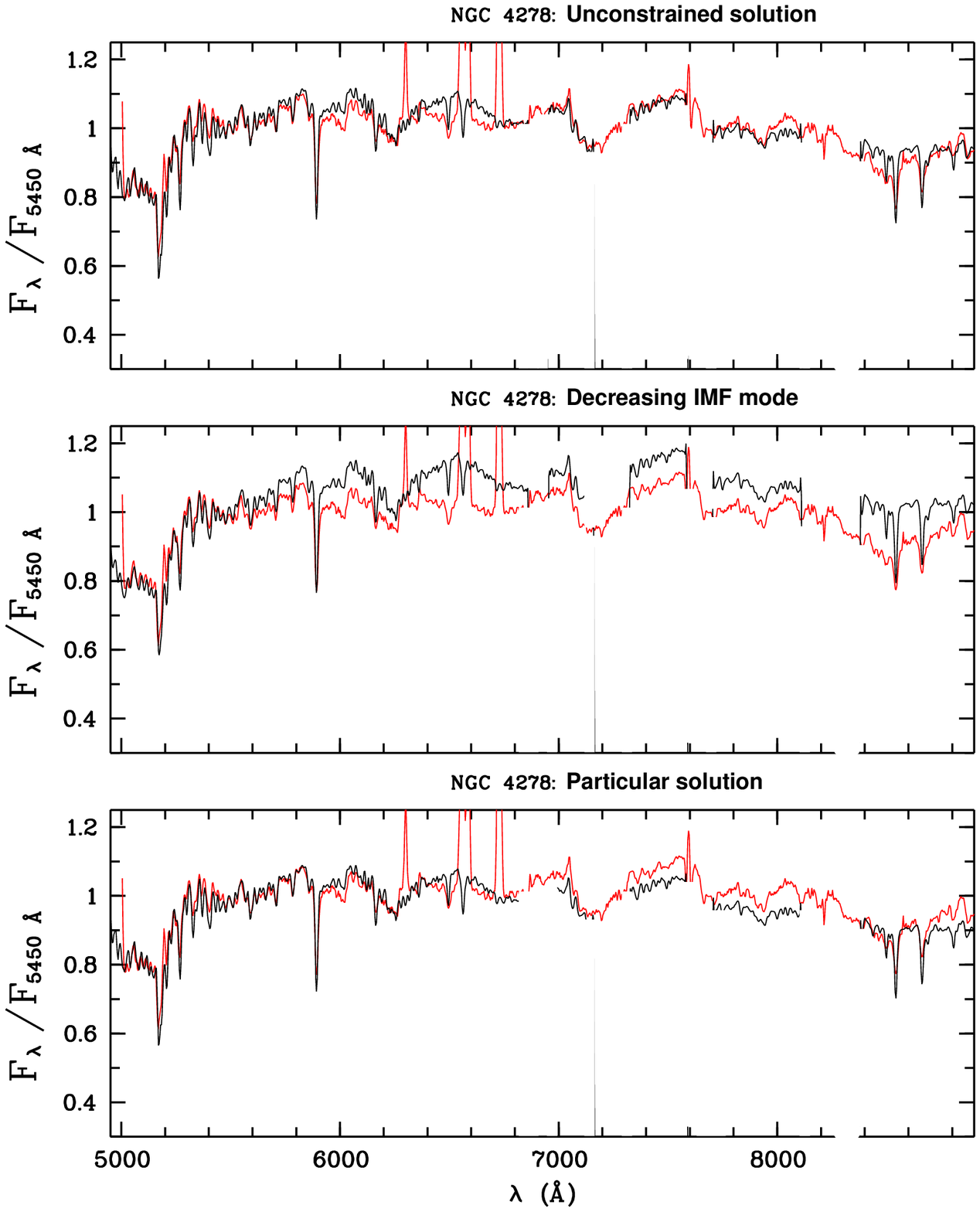}}}
\caption{NGC 4278: Same as Fig. \ref{G170a.eps}.}
\label{N4278.eps}
\end{figure*}

The solution of the ``Decreasing IMF mode'' presents a much larger synthetic distance (with values exceeding the one of the unconstrained solution by more than 5$\sigma$). This shows that this mode is very constraining for the galactic region, a fact that is confirmed through the large discrepancy between the observed and the synthetic spectrum in Fig. \ref{N4278.eps}.

The ``Standard mode'' provides a solution equivalent to the unconstrained one. It shows a small contribution of G4V stars badly determined suggesting an earlier location of the turnoff. This led us to impose a value to the contribution of this spectral class and to obtain an acceptable solution (shown in the last column of table \ref{NGC4278}) presenting an earlier turnoff and satisfying the constraints of this mode.

%\begin{figure*}
%\resizebox{18cm}{!}{{\includegraphics{HRn4278.eps}}}
%\caption{NGC 4278: HR diagram. Same as in Fig. \ref{HRG170_1.eps}. }
%\label{HRN4278.eps}
%\end{figure*}

%\begin{sidewaystable*}[htbp]
\begin{table*}
\small
\begin{center}
\begin{tabular}{|r|c|c|c|c|c|c|} %c|c|c|c|}
\hline
          & Unconstrained        & Mass interval & Dec. IMF   & Mass interval & Standard   & Particular \\ % Unconstrained            &
Star      & solution    & Dec. IMF mode &  mode    & Standard mode &  mode        & solution \\ % solution (2) & 
\hline
O7-BOV    & $0$          & $17 M_\odot-30 M_\odot$ & $0$         & $17 M_\odot-30 M_\odot$ & $0$        & $0$\\ %& $0\pm69$     
B3-4V     & $0$          & $ 3M_\odot-17 M_\odot$ & $0$         & $ 3M_\odot-17 M_\odot$ & $0$        & $0$\\ %& $0\pm58$     
A1-3V     & $0$          & $ 1.6M_\odot-3 M_\odot$ & $0$         & $ 1.6M_\odot-3 M_\odot$ & $0$        & $0$\\ %& $0\pm16$     
F2V       & $0$          & $ 0.8M_\odot-1.6 M_\odot$ &$0$         & $ 1.4M_\odot-1.6 M_\odot$ &$0$        & $0$\\ %& $0\pm35$     
F8-9V     & $0$          & $ 0.8M_\odot-1.6 M_\odot$ &$0$         & $ 1.1M_\odot-1.4 M_\odot$ &$0$        & $0$\\ %& $0\pm55$     
G4V       & $0$          & $ 0.8M_\odot-1.6 M_\odot$ &$1\pm2$     & $ 0.8M_\odot-1.1 M_\odot$ &$0.2\pm1$        & $\underline{7}$\\ %& $0\pm31$     
G9-K0V    & $0$          & $ \leq 0.8 M_\odot$ &$0$         & $  0.7M_\odot-0.8 M_\odot$ &$0$        & $0$\\ %& $0\pm24$     
K5V       & $0$          & $ \leq 0.8 M_\odot$ &$42\pm4$    & $  0.5M_\odot-0.7 M_\odot$ &$0$        & $0$\\ %& $0\pm40$     
M2V       & $0$          & $ \leq 0.8 M_\odot$ &$4\pm2$     & $  \leq 0.5 M_\odot$ &$0$        & $0$\\ %& $0\pm6$      
rG0IV     & $0$          & $ 0.8M_\odot-1.6 M_\odot$ &$0$         & $ 1.1M_\odot-1.4 M_\odot$ &$0$        & $0$ \\ %& $0\pm57$     
rG5IV     & $0$          & $ 0.8M_\odot-1.6 M_\odot$ &$0$         & $ 1.1M_\odot-1.4 M_\odot$ &$0$        & $0$\\ %& $0\pm24$     
rK0V      & $38\pm14$    & $ \leq 0.8 M_\odot$ &$0$         & $  0.7M_\odot-0.8 M_\odot$ &$37\pm15$  & $26\pm15$\\ %& $38\pm34$    
rK3V      & $38\pm8$     & $ \leq 0.8 M_\odot$ &$35\pm3$    & $  0.7M_\odot-0.8 M_\odot$ &$38\pm8$   & $40\pm5$\\ %& $38\pm27$    
rM1V      & $2\pm2$      & $ \leq 0.8 M_\odot$ &$5\pm4$     & $  \leq 0.5 M_\odot$ &$2.5\pm2$    & $3\pm2$\\ %& $2\pm6$      
G0-4III   & $0$          & $ 0.8M_\odot-1.6 M_\odot$ &$0$         & $ 1.1M_\odot-1.4 M_\odot$ &$0$        & $0$\\ %& $0\pm21$     
wG8III    & $0$          & $ 0.8M_\odot-1.6 M_\odot$ &$0$         & $ 1.1M_\odot-1.4 M_\odot$ &$0$        & $0$\\ %& $0\pm28$     
G9III     & $0$          & $ 0.8M_\odot-1.6 M_\odot$ &$0$         & $ 1.1M_\odot-1.4 M_\odot$ &$0$        & $0$\\ %& $0\pm30$     
K4III     & $0$          & $ 0.8M_\odot-1.6 M_\odot$ &$0$         & $ 1.1M_\odot-1.4 M_\odot$ &$0$        & $0$\\ %& $0\pm19$     
M0.5III   & $5\pm2$      & $ 0.8M_\odot-1.6 M_\odot$ &$0$         & $ 1.1M_\odot-1.4 M_\odot$ &$5\pm2$    & $4\pm1$\\ %& $5\pm4$      
M4III     & $3.5\pm0.1$  & $ 0.8M_\odot-1.6 M_\odot$ &$3\pm0.2$   & $ 1.4M_\odot-1.6 M_\odot$ &$3.5\pm0.1$& $3\pm0.3$\\ %& $3.5\pm2$  
M5III     & $0$          & $ 0.8M_\odot-1.6 M_\odot$ &$0$         & $ 1.4M_\odot-1.6 M_\odot$ &$0$        & $0$\\ %& $0\pm1.5$    
rG9III    & $0$          & $ 0.8M_\odot-1.6 M_\odot$ &$0$         & $ 1.1M_\odot-1.4 M_\odot$ &$0$        & $0$\\ %& $0\pm22$     
rK3III    & $3\pm6$      & $ 0.8M_\odot-1.6 M_\odot$ &$0$         & $ 1.1M_\odot-1.4 M_\odot$ &$3\pm7$    & $4\pm6$\\ %& $3\pm11$     
rK3IIIbis & $0$          & $ 0.8M_\odot-1.6 M_\odot$ &$0$         & $ 1.1M_\odot-1.4 M_\odot$ &$0$        & $0$\\ %& $0\pm17$     
rK5III    & $10\pm4$     & $ 0.8M_\odot-1.6 M_\odot$ &$10\pm2$    & $ 1.1M_\odot-1.4 M_\odot$ &$10\pm4$   & $11\pm4$\\ %& $10\pm8$     
G0Iab     & $0$          & $ 3M_\odot-17 M_\odot$ &$0$         & $ 3M_\odot-17 M_\odot$ &$0$        & $0$\\ %& $0\pm25$     
K4Iab     & $0$          & $ 3M_\odot-17 M_\odot$ & $0$         & $ 3M_\odot-17 M_\odot$ & $0$        & $0$\\ %& $0\pm19$     
M2Ia      & $0$          & $17 M_\odot-30 M_\odot$ & $0$         & $17 M_\odot-30 M_\odot$ & $0$        & $0$\\ %& $0\pm4$      
rG2Iab    & $0$          & $ 3M_\odot-17 M_\odot$ & $0$         & $ 3M_\odot-17 M_\odot$ & $0$        & $0$\\ %& $0\pm18$     
rK0II     & $0$          & $ 3M_\odot-17 M_\odot$ & $0$         & $ 3M_\odot-17 M_\odot$ & $0$        & $0$\\ %& $0\pm50$     
rK3Iab    & $0$          & $ 3M_\odot-17 M_\odot$ & $0$         & $ 3M_\odot-17 M_\odot$ & $0$        & $0$\\ %& $0\pm35$     
\hline
$D^2$ or $\chi^2$         & $12.3\pm0.95$     &  & $17.1\pm1.6$    &  & $12.3\pm0.9$     & $13.0\pm1.2$\\ %& $12.3\pm0.2$ \\
\hline
E(B-V)    & $0.02$       & & $0.00$ R    &  & $0.02$     & $0.00$\\ %& $0.02$ 
\hline
\end{tabular}
\end{center}
\caption{The different synthesis solutions for the nucleus of NGC4278. Same notations as in table \ref{G170}. The underlined contribution in the particular solution is imposed.} In the reddening row, R means that the synthetic spectrum is redder than the observed spectrum.
\label{NGC4278}
\end{table*}
%\end{sidewaystable*}

\subsection{The nucleus of the starburst galaxy NGC3310:}
According to the unconstrained solution, the MS stars and the metallic stars contribute, respectively, to $\sim94\%$ and $\sim68\%$ of the luminosity in the nucleus of NGC 3310 (table \ref{NGC3310}), implying a stellar population dominated by MS and metallic stars. The best defined turnoff is situated in A1-3V but a turnoff in O7-B0V is possible as well since a small contribution of these stars is present but badly determined. 
The reddening is very high in agreement with the location of the turnoff that indicates an important event of star formation and consequently the presence of a big quantity of dust in the region.

\begin{figure*}
\resizebox{15cm}{!}{{\includegraphics{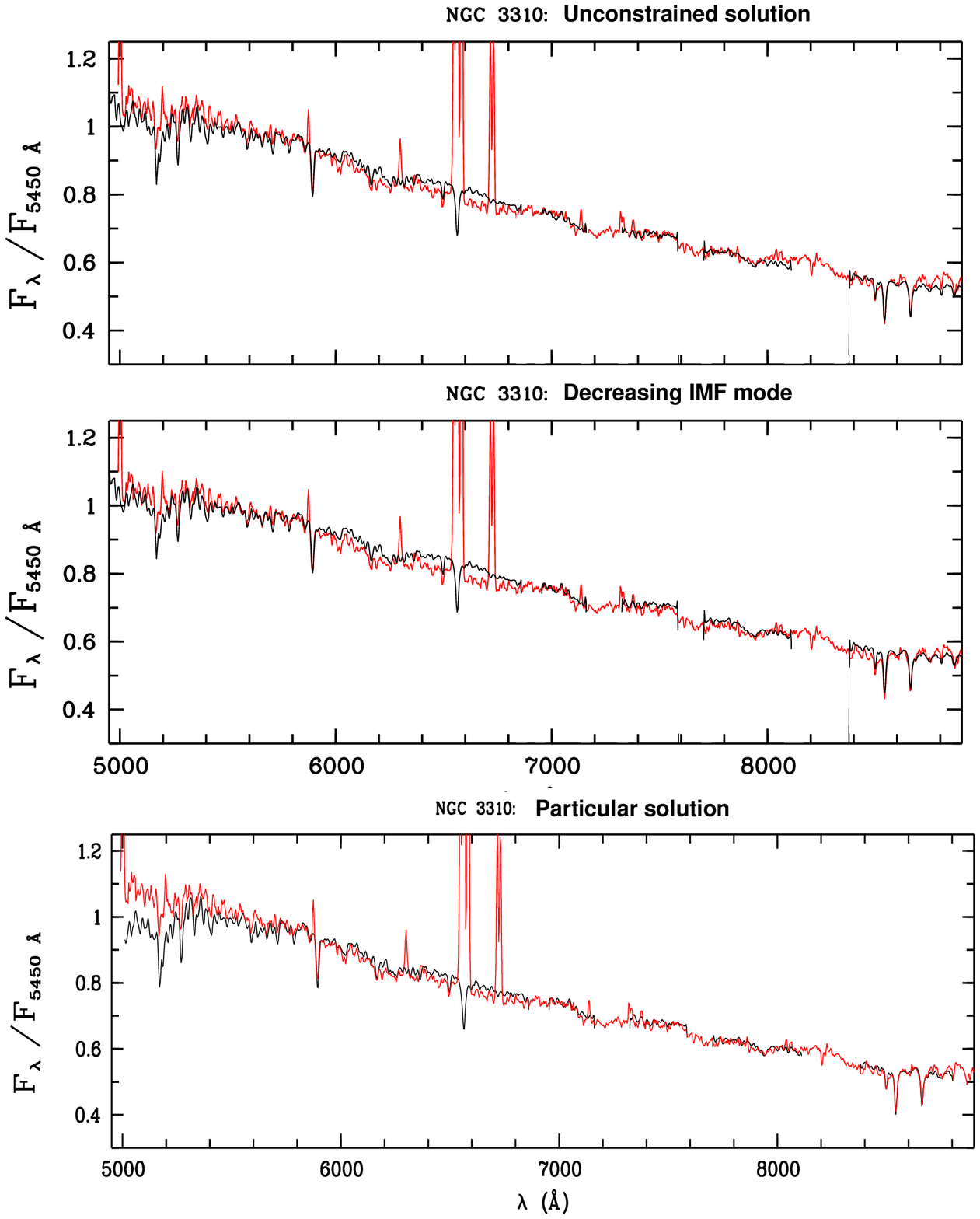}}}
\caption{NGC 3310: Same as Fig. \ref{G170a.eps}.}
\label{N3310.eps}
\end{figure*}

The solution of the ``Decreasing IMF'' mode presents a synthetic distance a little more than $\sim 1\sigma$ higher than the one of the unconstrained solution. It may therefore be acceptable (see also Fig. \ref{N3310.eps}). This solution distributes the non zero contributions to a larger number of dwarfs and confirm the high contribution to luminosity of dwarfs and metallic stars (resp. $\sim95\%$ and $\sim49\%$) as well as the location of the turnoff situated in A1-3 V (corresponding to an age of 200 million years for the last burst of star formation). In this solution, the constraint involving the O7-B0V stars is satisfied on the border of the domain of constraints (i.e. with equalities). This suggests that the ``Decreasing IMF mode''is very restricting in this object; the absence of the hot O7-B0V stars might therefore not be real. 
This conclusion is supported by the presence of emission lines in the spectrum of the starburst galaxy which suggests an ongoing star formation occuring in the nucleus of this galaxy.\\

The same scenario occurs in the ``Standard mode'' where the solution provides an acceptable synthetic distance. The best defined turnoff is situated in G5IV but small non zero contributions (not well defined) show that a turnoff at earlier types is possible as well. We show in the last column of table \ref{NGC3310} the example of such a situation where we impose the contributions of B2-3V star in addition to the ``Standard mode'' constraints. In this solution, the synthetic distance is acceptable which leads us to the same conclusion as previously.

%\begin{figure*}
%\resizebox{18cm}{!}{{\includegraphics{HRn3310.eps}}}
%\caption{NGC 3310: HR diagram. Same as in Fig. \ref{HRG170_1.eps}. }
%\label{HRn3310.eps}
%\end{figure*}

\begin{table*}
% \begin{sidewaystable}[htbp]
\small
\begin{center}
\begin{tabular}{|r|c|c|c|c|c|c|} %c|c|c|c|}
\hline
          & Unconstrained         & Mass interval & Dec. IMF   & Mass interval & Standard   & Particular\\ 
Etoile    & solution              & Dec. IMF mode &  mode    &Standard mode & mode     & solution \\ 
\hline
O7-BOV    & $4\pm5$      & $17 M_\odot-30 M_\odot$ & $0$         & $17 M_\odot-30 M_\odot$ & $4\pm33$   & $0$ \\
B3-4V     & $0$          & $ 3M_\odot-17 M_\odot$ & $0$         & $ 3M_\odot-17 M_\odot$ & $0.1\pm44.5$& $\underline{10}$\\
A1-3V     & $21\pm9$     & $ 1.6M_\odot-3 M_\odot$ & $30\pm12.5$ & $ 1.6M_\odot-3 M_\odot$ & $22\pm26  $& $15\pm9$\\
F2V       & $0$          & $ 0.8M_\odot-1.6 M_\odot$ &$1\pm17$    & $ 1.4M_\odot-1.6 M_\odot$ &$0$        & $0$\\
F8-9V     & $6\pm6$      & $ 0.8M_\odot-1.6 M_\odot$ &$0$         & $ 1.1M_\odot-1.4 M_\odot$ &$6\pm9$    & $6.5\pm6$\\
G4V       & $0$          & $ 0.8M_\odot-1.6 M_\odot$ &$6\pm11$    & $ 0.8M_\odot-1.1 M_\odot$ &$0$        & $0$\\
G9-K0V    & $0$          & $ \leq 0.8 M_\odot$ &$0$         & $  0.7M_\odot-0.8 M_\odot$ &$0$        & $0$\\
K5V       & $0$          & $ \leq 0.8 M_\odot$ &$11\pm4$    & $  0.5M_\odot-0.7 M_\odot$ &$0$        & $0$\\
M2V       & $0$          & $ \leq 0.8 M_\odot$ &$2\pm1$     & $  \leq 0.5 M_\odot$ &$0$        & $0$\\
rG0IV     & $0$          & $ 0.8M_\odot-1.6 M_\odot$ &$0$         & $ 1.1M_\odot-1.4 M_\odot$ &$0$        & $0$\\
rG5IV     & $30\pm10.5$  & $ 0.8M_\odot-1.6 M_\odot$ &$27\pm9$    & $ 1.1M_\odot-1.4 M_\odot$ &$30\pm10$  & $29\pm10$\\
rK0V      & $33\pm5.5$   & $ \leq 0.8 M_\odot$ &$18\pm7$    & $  0.7M_\odot-0.8 M_\odot$ &$33\pm4.5$ & $34\pm5$\\
rK3V      & $0$          & $ \leq 0.8 M_\odot$ &$0$         & $  0.7M_\odot-0.8 M_\odot$ &$0$        & $0$\\
rM1V      & $0$          & $ \leq 0.8 M_\odot$ &$0$         & $  \leq 0.5 M_\odot$ &$0$        & $0$\\
G0-4III   & $0$          & $ 0.8M_\odot-1.6 M_\odot$ &$0$         & $ 1.1M_\odot-1.4 M_\odot$ &$0$        & $0$\\
wG8III    & $0$          & $ 0.8M_\odot-1.6 M_\odot$ &$0$         & $ 1.1M_\odot-1.4 M_\odot$ &$0$        & $0$\\
G9III     & $0$          & $ 0.8M_\odot-1.6 M_\odot$ &$0$         & $ 1.1M_\odot-1.4 M_\odot$ &$0$        & $0$\\
K4III     & $0$          & $ 0.8M_\odot-1.6 M_\odot$ &$0$         & $ 1.1M_\odot-1.4 M_\odot$ &$0$        & $0$\\
M0.5III   & $0$          & $ 0.8M_\odot-1.6 M_\odot$ &$0$         & $ 1.1M_\odot-1.4 M_\odot$ &$0$        & $0$\\
M4III     & $1\pm0.2$    & $ 0.8M_\odot-1.6 M_\odot$ &$1\pm0.2$   & $ 1.4M_\odot-1.6 M_\odot$ &$1\pm0.2$  & $1\pm0.2$\\
M5III     & $0$          & $ 0.8M_\odot-1.6 M_\odot$ &$0$         & $ 1.4M_\odot-1.6 M_\odot$ &$0$        & $0$\\
rG9III    & $0$          & $ 0.8M_\odot-1.6 M_\odot$ &$0$         & $ 1.1M_\odot-1.4 M_\odot$ &$0$        & $0$\\
rK3III    & $0$          & $ 0.8M_\odot-1.6 M_\odot$ &$0$         & $ 1.1M_\odot-1.4 M_\odot$ &$0$        & $0$\\
rK3IIIbis & $0$          & $ 0.8M_\odot-1.6 M_\odot$ &$0$         & $ 1.1M_\odot-1.4 M_\odot$ &$0$        & $0$\\
rK5III    & $5\pm2$      & $ 0.8M_\odot-1.6 M_\odot$ &$4\pm2$     & $ 1.1M_\odot-1.4 M_\odot$ &$5\pm2$    & $5\pm2$\\
G0Iab     & $0$          & $ 3M_\odot-17 M_\odot$ &$0$         & $ 3M_\odot-17 M_\odot$ &$0$        & $0$\\
K4Iab     & $0$          & $ 3M_\odot-17 M_\odot$ & $0$         & $ 3M_\odot-17 M_\odot$ & $0$        & $0$\\
M2Ia      & $0$          & $17 M_\odot-30 M_\odot$ & $0$         & $17 M_\odot-30 M_\odot$ & $0$        & $0$\\
rG2Iab    & $0$          & $ 3M_\odot-17 M_\odot$ & $0$         & $ 3M_\odot-17 M_\odot$ & $0$        & $0$\\
rK0II     & $0$          & $ 3M_\odot-17 M_\odot$ & $0$         & $ 3M_\odot-17 M_\odot$ & $0$        & $0$\\
rK3Iab    & $0$          & $ 3M_\odot-17 M_\odot$ & $0$         & $ 3M_\odot-17 M_\odot$ & $0$        & $0$\\
\hline
$D^2$ or $\chi^2$        & $11.4\pm1.5$      & & $13.2\pm1.5$     & & $11.4\pm1.2$     & $11.6\pm1.5$\\
\hline
E(B-V)    & $0.23$     &  & $0.2$     &  & $0.23$     & $0.23$\\
\hline
\end{tabular}
\end{center}
\caption{Results of the spectral synthesis of NGC 3310 nucleus. Same notations as in tables \ref{G170} and \ref{NGC4278}.}
\label{NGC3310}
\end{table*}

\subsection{The nucleus of the Seyfert2 galaxy NGC2110:}
The population in the unconstrained solution is dominated by dwarf stars and is moderately metallic ($\sim68\%$ of the optical luminosity is due to MS stars and $\sim52\%$ is due to overabundant stars, see table \ref{NGC2110} and  Fig. \ref{NGC2110bis.eps}). The best defined turnoff is situated in K0V but according to the previous discussion, an earlier turnoff is possible as well.
% especially, the presence of O7-B0V is quite likely if the nebular emission lines due to hot stars are fainter than the observed ones (due mainly to the AGN).\\

The solution of the ``Decreasing IMF'' mode is acceptable since its synthetic distance is slightly higher than the one of the unconstrained solution and lies at less than $1 \sigma$ from this one. In this mode, the contribution of dwarf stars is enhanced ($\sim78\%$) while overabundant stars contribute less to the visible luminosity (only $\sim38\%$). The best defined turnoff is situated in K3V but a turnoff in F2V is also possible, a fact that is confirmed in the particular solution where we imposed a contribution of $\sim3\%$ to the star class F2V (see also Fig. \ref{NGC2110bis.eps}).  

In the ``Standard mode'', the solution presents a synthetic distance equal to the one of the unconstrained solution. The solution is very similar to the unconstrained one.
   
\begin{figure*}
\resizebox{15cm}{!}{{\includegraphics{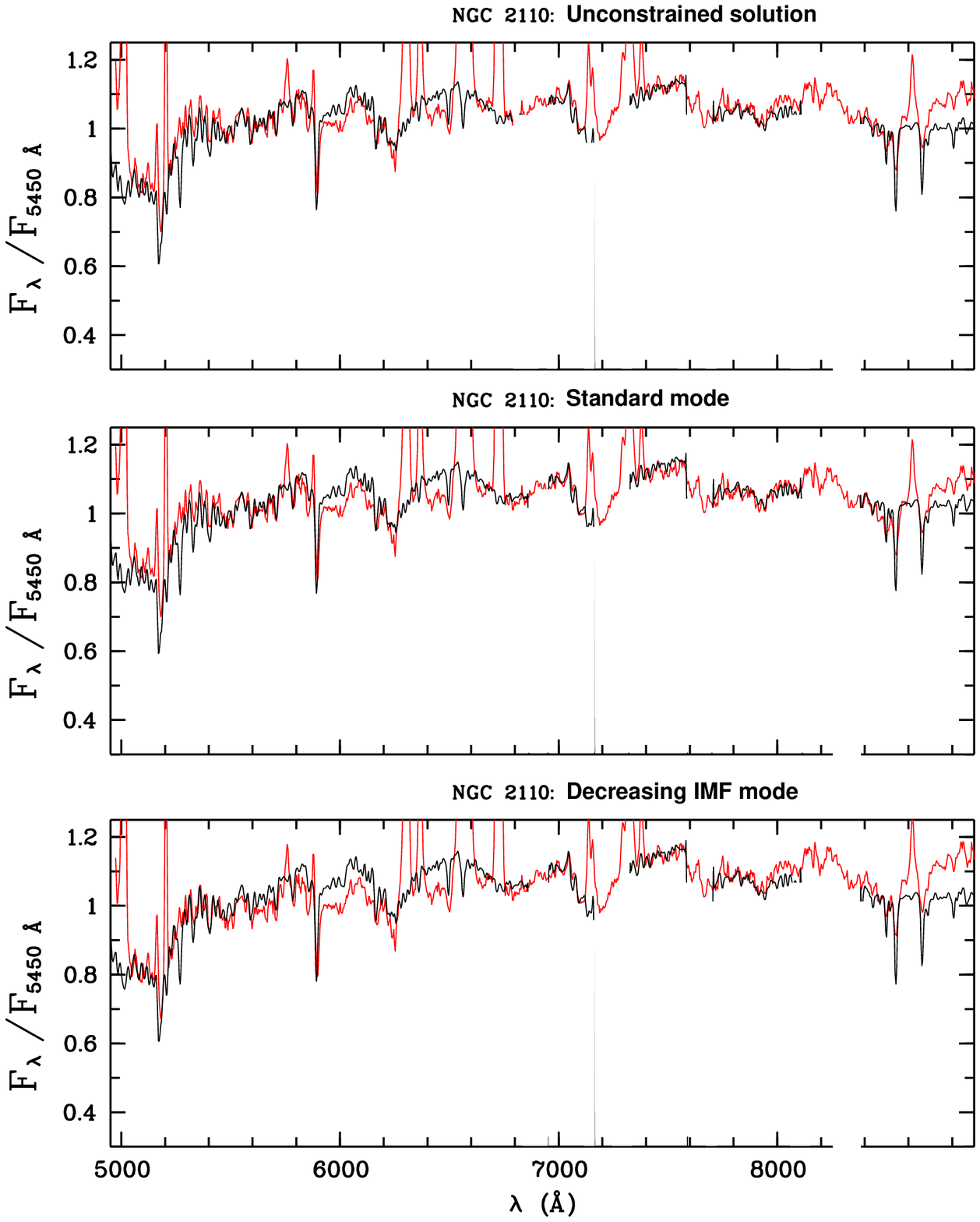}}}
\caption{NGC 2110: Same as Fig. \ref{G170a.eps}.}
\label{N2110.eps}
\end{figure*}

\begin{figure*}
\resizebox{15cm}{!}{{\includegraphics{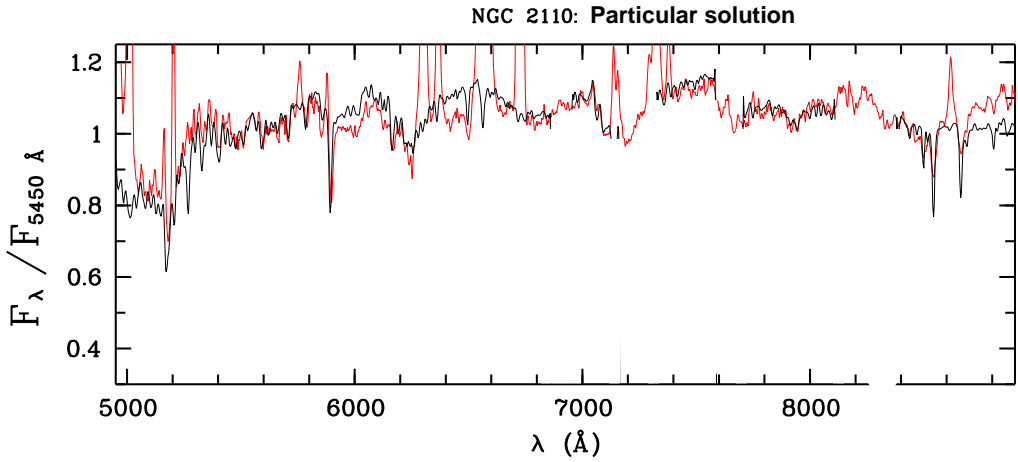}}}
\caption{NGC 2110: Particular solution. Same as Fig. \ref{G170a.eps}. }
\label{NGC2110bis.eps}
\end{figure*}

%\begin{figure*}
%\resizebox{18cm}{!}{{\includegraphics{HRn2110_1.eps}}}
%\caption{NGC 2110: HR diagram. Same as in Fig. \ref{HRG170_1.eps}. }
%\label{HRn2110_1.eps}
%\end{figure*}

%\begin{figure*}
%\resizebox{12cm}{!}{{\includegraphics{HRn2110_2bis.eps}}}
%\caption{NGC 2110: HR diagram. Same as in Fig. \ref{HRG170_1.eps}. }
%\label{HRn2110_2.eps}
%\end{figure*}

%\begin{sidewaystable*}[htbp]
\begin{table*}
\small
\begin{center}
\begin{tabular}{|r|c|c|c|c|c|c|} %c|c|c|c|c|}
\hline
          & Unconstrained         & Mass interval & Dec. IMF   & Mass interval & Standard   & Particular  \\ %& Imposed \\ 
Etoile    & solution  & Dec. IMF mode &  mode   &Standard mode & mode    &  solution  \\ %& contributions (2)\\ 
\hline
O7-BOV    & $3\pm7$      & $17 M_\odot-30 M_\odot$ & $0$        & $17 M_\odot-30 M_\odot$ & $2\pm67$   & $0$       \\ %& $\underline{5}$\\
B3-4V     & $0$          & $ 3M_\odot-17 M_\odot$ & $0$        & $ 3M_\odot-17 M_\odot$ & $0.1\pm71$ & $0$       \\ %& $\underline{7}$\\
A1-3V     & $0$          & $ 1.6M_\odot-3 M_\odot$ & $0$        & $ 1.6M_\odot-3 M_\odot$ & $0$        & $0$       \\ %& $\underline{1}$\\
F2V       & $0$          & $ 0.8M_\odot-1.6 M_\odot$ &$0.006\pm7$        & $ 1.4M_\odot-1.6 M_\odot$ &$0$        & $\underline{3}$ \\ %& $0$\\
F8-9V     & $0$          & $ 0.8M_\odot-1.6 M_\odot$ &$0$        & $ 1.1M_\odot-1.4 M_\odot$ &$0$        & $0$       \\ %& $0$\\
G4V       & $0$          & $ 0.8M_\odot-1.6 M_\odot$ &$3\pm18$   & $ 0.8M_\odot-1.1 M_\odot$ &$0$        & $2\pm14$  \\ %& $\underline{1}$\\
G9-K0V    & $0$          & $ \leq 0.8 M_\odot$ &$0$        & $  0.7M_\odot-0.8 M_\odot$ &$0$        & $0$       \\ %& $0$\\
K5V       & $9\pm8$      & $ \leq 0.8 M_\odot$ &$32\pm6.5$   & $  0.5M_\odot-0.7 M_\odot$ &$9\pm10$   & $32\pm7$  \\ %& $2\pm6$\\
M2V       & $6\pm2$      & $ \leq 0.8 M_\odot$ &$6\pm1$    & $  \leq 0.5 M_\odot$ &$6\pm1$    & $6\pm1$   \\ %& $9\pm1$\\
rG0IV     & $0$          & $ 0.8M_\odot-1.6 M_\odot$ &$0$        & $ 1.1M_\odot-1.4 M_\odot$ &$0$        & $0$       \\ %& $0$\\
rG5IV     & $9\pm12$     & $ 0.8M_\odot-1.6 M_\odot$ &$10\pm16$  & $ 1.1M_\odot-1.4 M_\odot$ &$10\pm12$  & $9\pm15$  \\ %& $0$\\
rK0V      & $19\pm6$     & $ \leq 0.8 M_\odot$ &$0$        & $  0.7M_\odot-0.8 M_\odot$ &$19\pm8$   & $0$       \\ %& $24\pm0.4$\\
rK3V      & $22\pm5$     & $ \leq 0.8 M_\odot$ &$27\pm3$   & $  0.7M_\odot-0.8 M_\odot$ &$22\pm7$   & $26\pm2$  \\ %& $20.5\pm4$\\
rM1V      & $0$          & $ \leq 0.8 M_\odot$ &$0$        & $  \leq 0.5 M_\odot$ &$0$        & $0$       \\ %& $0$\\
G0-4III   & $0$          & $ 0.8M_\odot-1.6 M_\odot$ &$0$        & $ 1.1M_\odot-1.4 M_\odot$ &$0$        & $0$       \\ %& $0$\\
wG8III    & $0$          & $ 0.8M_\odot-1.6 M_\odot$ &$0$        & $ 1.1M_\odot-1.4 M_\odot$ &$0$        & $0$       \\ %& $0$\\
G9III     & $0$          & $ 0.8M_\odot-1.6 M_\odot$ &$0$        & $ 1.1M_\odot-1.4 M_\odot$ &$0$        & $0$       \\ %& $0$\\
K4III     & $15\pm5$     & $ 0.8M_\odot-1.6 M_\odot$ &$6\pm4$    & $ 1.1M_\odot-1.4 M_\odot$ &$15\pm7$   & $6\pm4$   \\ %& $13\pm6$\\
M0.5III   & $12\pm3$     & $ 0.8M_\odot-1.6 M_\odot$ &$12\pm3$   & $ 1.1M_\odot-1.4 M_\odot$ &$12\pm1.5$ & $11.5\pm3$\\ %& $7\pm3$\\
M4III     & $2\pm0.3$    & $ 0.8M_\odot-1.6 M_\odot$ &$2\pm0.2$  & $ 1.4M_\odot-1.6 M_\odot$ &$2\pm0.1$  & $2\pm0.2$ \\ %& $2\pm0.5$\\
M5III     & $0$          & $ 0.8M_\odot-1.6 M_\odot$ &$0$        & $ 1.4M_\odot-1.6 M_\odot$ &$0$        & $0$     \\ %& $0.1\pm0.5$\\
rG9III    & $0$          & $ 0.8M_\odot-1.6 M_\odot$ &$0$        & $ 1.1M_\odot-1.4 M_\odot$ &$0$        & $0$       \\ %& $0$\\
rK3III    & $0$          & $ 0.8M_\odot-1.6 M_\odot$ &$0$        & $ 1.1M_\odot-1.4 M_\odot$ &$0$        & $0$       \\ %& $0$\\
rK3IIIbis & $0$          & $ 0.8M_\odot-1.6 M_\odot$ &$0$        & $ 1.1M_\odot-1.4 M_\odot$ &$0$        & $0$       \\ %& $0$\\
rK5III    & $2.5\pm7$    & $ 0.8M_\odot-1.6 M_\odot$ &$1\pm8$    & $ 1.1M_\odot-1.4 M_\odot$ &$2.5\pm4$  & $2\pm8$   \\ %& $8\pm6$\\
G0Iab     & $0$          & $ 3M_\odot-17 M_\odot$ &$0$        & $ 3M_\odot-17 M_\odot$ &$0$        & $0$       \\ %& $0$\\
K4Iab     & $0$          & $ 3M_\odot-17 M_\odot$ & $0$        & $ 3M_\odot-17 M_\odot$ & $0$        & $0$       \\ %& $0$\\
M2Ia      & $0$          & $17 M_\odot-30 M_\odot$ & $0$        & $17 M_\odot-30 M_\odot$ & $0$        & $0$       \\ %& $0.2\pm2$\\
rG2Iab    & $0$          & $ 3M_\odot-17 M_\odot$ & $0$        & $ 3M_\odot-17 M_\odot$ & $0$        & $0$       \\ %& $0$\\
rK0II     & $0$          & $ 3M_\odot-17 M_\odot$ & $0$        & $ 3M_\odot-17 M_\odot$ & $0$        & $0$       \\ %& $0$\\
rK3Iab    & $0$          & $ 3M_\odot-17 M_\odot$ & $0$        & $ 3M_\odot-17 M_\odot$ & $0$        & $0$       \\ %& $0$\\
\hline
$D^2$ or $\chi^2$       & $13.1\pm 2.0$     &  & $13.5\pm1.9$   &  & $13.1\pm1.2$     & $13.5\pm1.9$       \\ %& $13.8\pm2.1$\\
\hline
E(B-V)    & $0.05$     &  & $0.0$     & & $0.05$     & $0.05$       \\ %& $0.08$\\
\hline
\end{tabular}
\end{center}
\caption{Results of the spectral synthesis of NGC 2110 nucleus. Same notations as in tables \ref{G170} and \ref{NGC4278}.}
\label{NGC2110}
\end{table*}
%\end{sidewaystable*}

\section{Conclusion}\label{sec:concl}

The ideal case for a spectral synthesis giving a synthetic distance 
equal to zero would be the case where the signal to noise ratio of
the galactic and stellar spectra goes to infinity and where the stellar
database is complete. In such a case, all stars with spectral types
later than the spectral type at the turnoff position would have non
zero contributions to luminosity. But in practice all unconstrained
solutions show many zero contributions; this is due to the finite
signal to noise ratio of our spectra and to the limitation of the
stellar database which itself is due to the finite spectral
resolution.

Therefore, constraining the stellar population would {\it {a priori}}
reduce the number of zero contributions because of the additive
information introduced in this process. But as can be seen in the
previous results, no large improvement in eliminating the zero
contributions has appeared. This is probably due to the not perfect adequation of the observational data. Actually  as the
constraints are expressed by large inequalities (i.e. equalities are
allowed) optima are usually located on the border of the
domain in which solutions are constrained.\\

The stellar synthesis method with constraints presented in this paper has been applied to the 27 regions of galaxies studied in Paper IV.
In general, all 27 regions present ``Standard mode'' solutions equal or very
similar to the unconstrained solution. Moreover, all zero contributions in the
unconstrained solutions remain null in the ``Standard mode'' or have
small ill-defined values and all well- and ill- determined
contributions remain respectively well- and ill- defined. This result 
shows that ``Standard mode'' solutions
are generally included inside the error bars of the unconstrained
solution and when they are not, their synthetic distances are at
several $\sigma$ from that of the unconstrained solution. \\

In the ``Decreasing IMF mode', the number of star classes contributing
to the synthesis is often larger than that of the unconstrained
solution and of the ``Standard mode''. This fact affects especially
dwarf stars and is due to the sharper distribution of stars in the
mass groups of the H-R diagram in this mode. For the same reason and
because the number of constraints is larger, the synthetic distances
here are in general larger than those of the previous mode and of the
unconstrained solution.\\

In both modes (``Standard'' and ``Decreasing IMF'') some solutions
satisfy their constraints on the border of the domain. This shows that
in such cases constraints are somehow too strong and induce bias.
However, these solutions provide some indications, thanks to the error
bars, allowing one to find acceptable solutions that satisfy the
desired conditions inside the domain of constraints (see previous
examples). \\

 As a matter of fact, the solution of the least square problem is the
one that minimizes the synthetic distance; this happens often on the
border of the domain of constraints but the goal is not to find the
optimal mathematical solution, rather a ``realistic'' or physical one
next to the minimum. \\

All previous results are very well confirmed in the case of the
globular cluster G170. This is a very important point since this
object is constituted of a single burst of star formation;
consequently, any deviation of the behaviour of the resulting stellar
population due to the inclusion of astrophysical constraints can
clearly be detected in this object. \\

This study has shown that the inverse method described in this paper
and in Papers I, II and III is very stable against the inclusion
of additional astrophysical constraints, and is, therefore, very
reliable.\\

However, constraining the solutions and using the information provided
by the error analysis allows one to find similar solutions with
younger bursts of star formation. Thus, it is crucial to perform tests
such as in the previous section and to discuss the results, especially
the different possible locations of the turnoffs, i.e. the age of the
last burst of star formation.\\

\appendix 
\section{Calculus of the standard deviation on the synthetic distance:}
In this appendix, we compute the standard deviations on the synthetic distance due to observational errors around the studied object. This calculation is complementary to the error analysis made in Paper III where only the standard deviations around the stellar contributions and the variance-covariance matrices were computed. As the synthetic distance is a scalar, its variance-covariance matrix is reduced to its variance.\\
Thus, here we search for the deviations $d(D^2)$ around the square of the obtained synthetic distance $D_0^2$ due to deviations $d\bm W_{obs}$ around the observation $\bm W_{{obs\,0}}$. We recall that this computation is only valuable in the overdetermined case (in the underdetermined case, this distance is equal to zero). \\

If we make a change of variables on the equivalent widths as $W'_j=P_j^{1/2} W_j$, the square of the synthetic distance will be written as follows:  
\begin{equation}
D^2=(\bm{W'}_{syn}-\bm{W'}_{obs})^T(\bm{W'}_{syn}-\bm
W'_{obs})
\end{equation}
Then a differenciation around $D_0^2$ gives:
\begin{equation}
dD^2=2(d\bm W'_{syn}-d\bm W'_{obs})^T(\bm{W'}_{syn\,0}-\bm{W'}_{obs\,0})\,.
\end{equation}
Now, if we replace $d\bm W'_{syn}$ by $\bM H d\bm W'_{obs}$, where $\bM H$ is the orthogonal projector on the synthetic surface (see Paper III), then we get:
\begin{equation}
\begin{array}{ll}
dD^2&=2((\bM H-\bM{Id})d\bm W'_{obs})^T(\bm{W'}_{syn\,0}-\bm{W'}_{obs\,0})\\
     &=2d\bm {W'}_{obs}^T(\bM H^T-\bM{Id})(\bm{W'}_{syn\,0}-\bm{W'}_{obs\,0})\,.
\end{array}
\end{equation}
where $\bM{Id}$ is the identity matrix.\\
Let us write, on the one hand, the definition of the variance of the quantity $dD^2$:
\begin{equation}
var(dD^2)=<dD^2dD^{2T}>-<dD^2>^2 \label{vard}
\end{equation}
on the other hand, we have $<dD^2>=2D_0<dD>$. Then if we translate the origin of the vector space in $\bm{W'}_{{syn 0}}$, we can consider the subspace of dimension 1 of which the generator vector is $\bm{W'}_{{obs\,0}}-\bm{W'}_{{syn\,0}}$. In this subspace the synthetic distance is described by the same vector $\bm D_0=\bm{W'}_{{obs\,0}}-\bm{W'}_{{syn\,0}}$ and the deviation to this distance is vector $d\bm D=\bm D-\bm D_0$. Thus, if we call $\bm u$ the unit vector of this subspace ($\bm u=\frac{\bm{W'}_{{obs\,0}}-\bm{W'}_{{syn\,0}}}{\|{\bm{W'}_{{obs\,0}}-\bm{W'}_{{syn\,0}}\|}}$), we can construct the orthogonal projector over it as $\bM P=\bm u\bm u^T$; then we get $d\bm D=\bM Pd\bm{W'}_{obs}$ and $<dD>=<d\bm D>=\bM P<d\bm{W'}_{syn}>=0$. This implies that $<dD^2>=0$ and $var(dD^2)=var(D^2)$.\\

Back to equation (\ref{vard}), we have:
\begin{equation}
\begin{array}{ll}
var(dD^2)&=var(D^2)\\
          &=<dD^2dD^{2}>\\
         &=4 \Delta \bm{W'}^T(\bM{H}-\bM{Id})<d\bm{W'}_{obs}d\bm{W'}_{obs}^T>(\bM{H}^T-\bM{Id})\Delta \bm{W'}
\end{array}
\end{equation}
where $\Delta \bm{W'}=\bm{W'}_{{obs\,0}}-\bm{W'}_{{syn\,0}}$. Then

\begin{equation}
\fbox{$
\sigma_{D^2}=2\sqrt{\Delta \bm{W'}^T(\bM{H}-\bM{Id})\bM
{V'}_{obs} (\bM{H}^T-\bM{Id})\Delta \bm{W'}}$}
\end{equation}
In addition, as $<dD>=<dD^2>=0$, we set the simple equation
\begin{equation}
\fbox{$\sigma_D=\frac{\sigma_{D^2}}{2D_0}.$}
\end{equation}

\begin{acknowledgements}
We would like to thank Pascale Jablonka for kindly providing the spectrum of the globular cluster G170.
\end{acknowledgements}

\end{document}